# Linear magnetoresistance and surface to bulk coupling in topological insulator thin films

**Sourabh Singh[1#], R.K. Gopal[1#], Jit Sarkar[1], Atul Pandey[1], Bhavesh G. Patel[2] and Chiranjib Mitra[1]**

*[1]Indian Institute of Science Education and Research Kolkata, Mohanpur, 741246*

*India*

*[2]Institute for Plasma Research, Bhat, Gandhinagar, 382428, India*

**ABSTRACT**:

We explore the temperature dependent magnetoresistance of bulk insulating topological insulator thin films. Thin films of $Bi_2Se_2Te$ and $BiSbTeSe_{1.6}$ were grown using Pulsed Laser Deposition technique and subjected to transport measurements. Magnetotransport measurements indicate a non-saturating linear magnetoresistance (LMR) behavior at high magnetic field values. We present a careful analysis to explain the origin of LMR taking into consideration all the existing models of LMR. Here we consider that the bulk insulating states and the metallic surface states constitute two parallel conduction channels. Invoking this we were able to explain linear magnetoresistance behavior as a competition between these parallel channels. We observe that the cross-over field, where LMR sets in, decreases with increasing temperature. We propose that this cross over field can be used phenomenologically to estimate the strength of surface to bulk coupling.

**INTRODUCTION**:

Topological insulators (TI) are a new state of quantum matter possessing extraordinary physical properties due to its linearly dispersed Dirac surface states residing in the bulk band gap. The unique arrangement of these surface states in the bulk band gap sets them apart from the trivial surface states such as Rashba spin split surface states[1,2]. The mechanism behind the appearance of these states is the bulk band inversion due to strong spin-orbit coupling. These classes of materials possess massless helical surface states which have topological protection against the non-magnetic disorder due to the presence of time reversal symmetry and a π Berry phase[3]. The helical nature of the surface states guarantees the absence of the backscattering from disorder as opposed to the trivial two-dimensional electron gas systems which localize the electrons in the presence of disorder at low temperatures.

An ideal TI consists of metallic surface states that are immune to non-magnetic disorder and a perfectly insulating bulk. But in real samples, the bulk carriers play an inhibiting role in the performance of TI samples[4]. Unintentional defect formation during sample growth results in undesirable bulk conduction which in turn masks the non-trivial contribution of the surface states. It is essential for a TI device to have a minimal bulk contribution to the conduction and in this regard $Bi_2Se_2Te$ (BST) is a worthy candidate[5–7]. The resistance vs. temperature (R-T) profile of these samples yields a bulk insulating response (see fig. 1(a)) thus enunciating that the chemical potential lies in the bulk band gap. This condition is known as purely topological transport and is a must for all TI based devices.

Magnetoresistance (MR) measurements yield a great deal of knowledge about the topological nature of surface states[8–12]. Shubnikov-de Hass oscillations (SdH)[9,13], Universal Conductance Fluctuations (UCF) and Weak antilocalization (WAL) have been observed in TI samples[14,15]. These phenomena reveal a plethora of information about the intricacies of TI and have been studied in detail. Along with the afore mentioned phenomena a TI also exhibits a non-saturating linear magnetoresistance (LMR) behavior at high magnetic field values[16–19]. Linear MR was first studied by Kapitza in detail in the year 1930[20]. He observed linear MR in metals with open Fermi surfaces. Understanding this LMR behavior is not only an important theoretical quest but also beneficial from an application point of view. This non-saturating LMR behavior can be used in magnetic sensors[21]. There has been a lot of debate on the origin of LMR in TI with different groups having prescribed different sources of origin of LMR[22]. The motivation of the present work is to understand the origin of LMR in bulk insulating samples which are in the topological transport regime. Systematic temperature dependent MR measurements provide deeper insights into the origin of LMR and the role of surface to bulk coupling in the onset of LMR.

## EXPERIMENTAL:

Thin films were deposited by Pulsed Laser Deposition (PLD) technique. KrF Excimer Laser ($\lambda$ = 248 nm) was used to ablate the BST and BSTS target. The deposition was done in a flowing Argon environment. The thicknesses of the films are about 200nm. The supplementary section is included regarding the details of thin film growth, characterization and measurement technique used in this work.

## RESULTS & DISCUSSION:

The Resistance vs. Temperature behavior(R-T) of BST sample exhibits an insulating behavior down to around 40K [Figure **1(a)**]. The R-T profile can be broadly divided into two regimes: thermally activated bulk carriers dominated regime at higher temperatures and metallic behavior of the surface states dominated regime below 40K.The presence of two components thus allows us to suitably map the R-T of TI with a parallel channel conduction model. In an earlier work Gao et al., carried out a gate dependent study of the magnetoresistance[2323]. They observed that for $V_g$ (gate voltage) < -40V the MR shows a parabolic behavior and the R-T profile shows a metallic character.  From this observation, it is evident that there exists an underlying relationship between the nature of R-T curves and the corresponding magnetoresistance at a particular temperature.

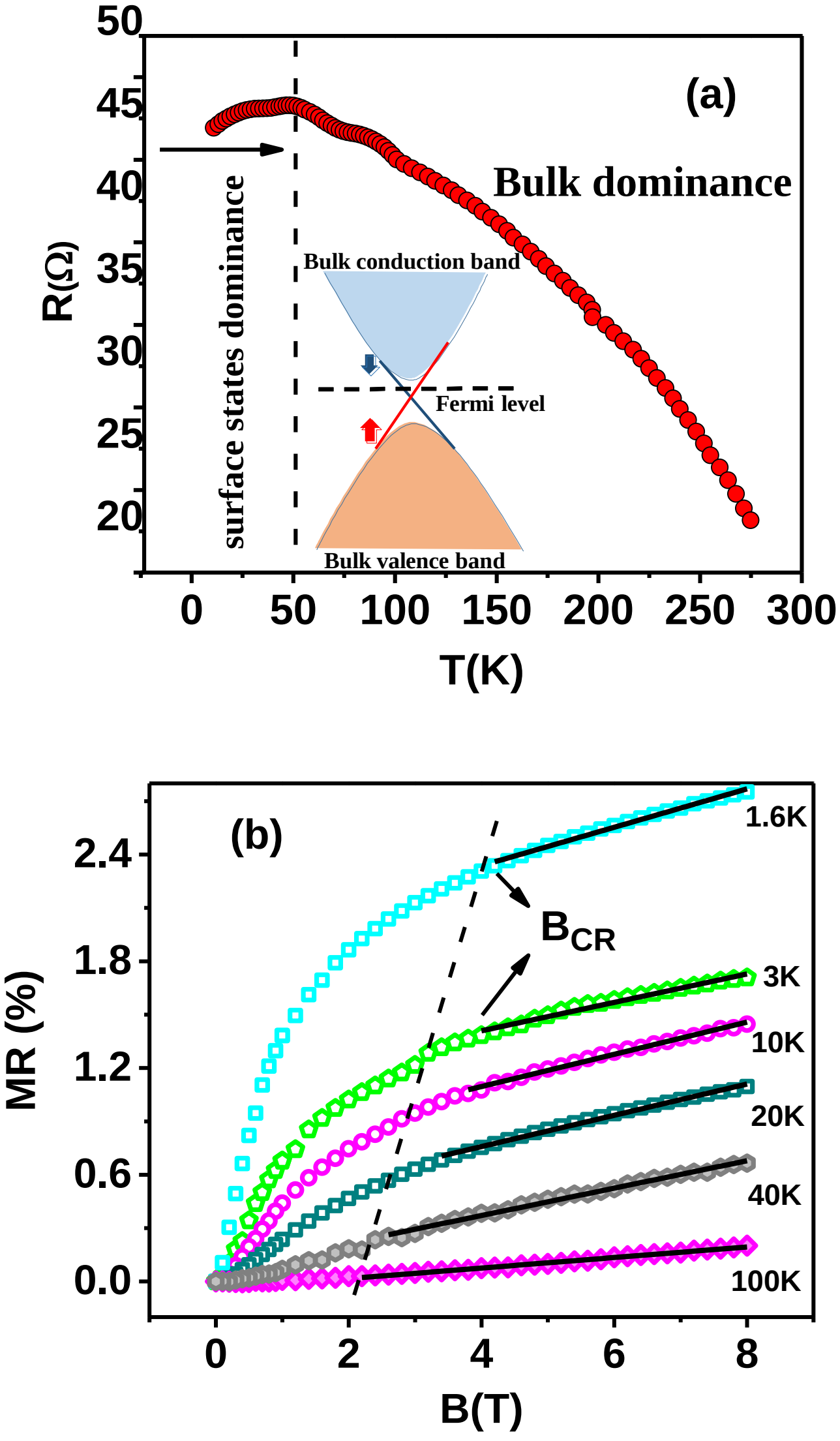

*FIG. **1**(a) Temperature dependent resistance of BST thin films. The temperature profile can be divided into two regimes which correspond to bulk dominating and surface states dominating regimes demarcated by the dashed line.(b) MR of BST sample at various temperatures the black solid line depicts the linear magneto-resistance regime.*

Figure 1(b) shows MR of BST thin film at different temperatures. MR shows a logarithmic cusp around zero magnetic field due to WAL[14,24] which is captured by the Hikami Larkin Nagaoka (HLN) equation. Increasing the magnetic field results in a nonsaturating linear response. This is shown in the figure by the solid black line. The dashed line depicts the onset of LMR and has a decreasing trend with increasing temperature. This type of unconventional LMR behavior has been observed earlier in metals with open fermi surfaces, narrow gap semiconductors or semi-metals[22]. Above 40K we can observe that the bulk conduction dominates over the metallic surface contribution (figure 1(a)). This is reflected in the bulk quadratic behavior seen in the MR plots in figure 1(b) (in the temperature above 40 K).

For conventional systems the application of a perpendicular magnetic field results in a change of resistance which is represented in the following form:

$$\frac{R(B,T)}{R(0,T)} = f\left[\frac{B}{R(0,T)}\right] \qquad (1)$$

Where R (0,T) is the resistance in zero magnetic field and f is a function which depends only on the geometrical configuration and on the sample type. This scaling of the magnetic field by the zero field resistance enables one to correlate the MR measurements performed at different temperatures. This is known as Kohler's rule and the corresponding plots are called Kohler's plots[25]. Kohler's rule stems from a simple argument that the product of the cyclotron frequency and the scattering time dictates the temperature dependent MR. The cyclotron frequency depends on the magnetic field and is independent of temperature whereas the scattering time is a temperature dependent term. Figure 2(a) and (b) show the Kohler's plot for sample BST and BSTS are shown respectively. If the sample were a simple system with a unique scattering time the MR measurements belonging to different temperatures would fall in the same curve. Thus this violation of Kohler's rule implies that LMR in TIs have a complex origin and the objective of the present work is to unravel this mystery. Earlier Abrikosov's quantum LMR[26,27] and Parish-Littlewood (PL)[28,29] model had been used to explain the origin of LMR in TIs[22,30,31]. The present MR data was analyzed and compared with the predictions of both the model; neither of Quantum LMR nor PL model interprets the origin of LMR in the present samples (c.f. Supplementary Section).

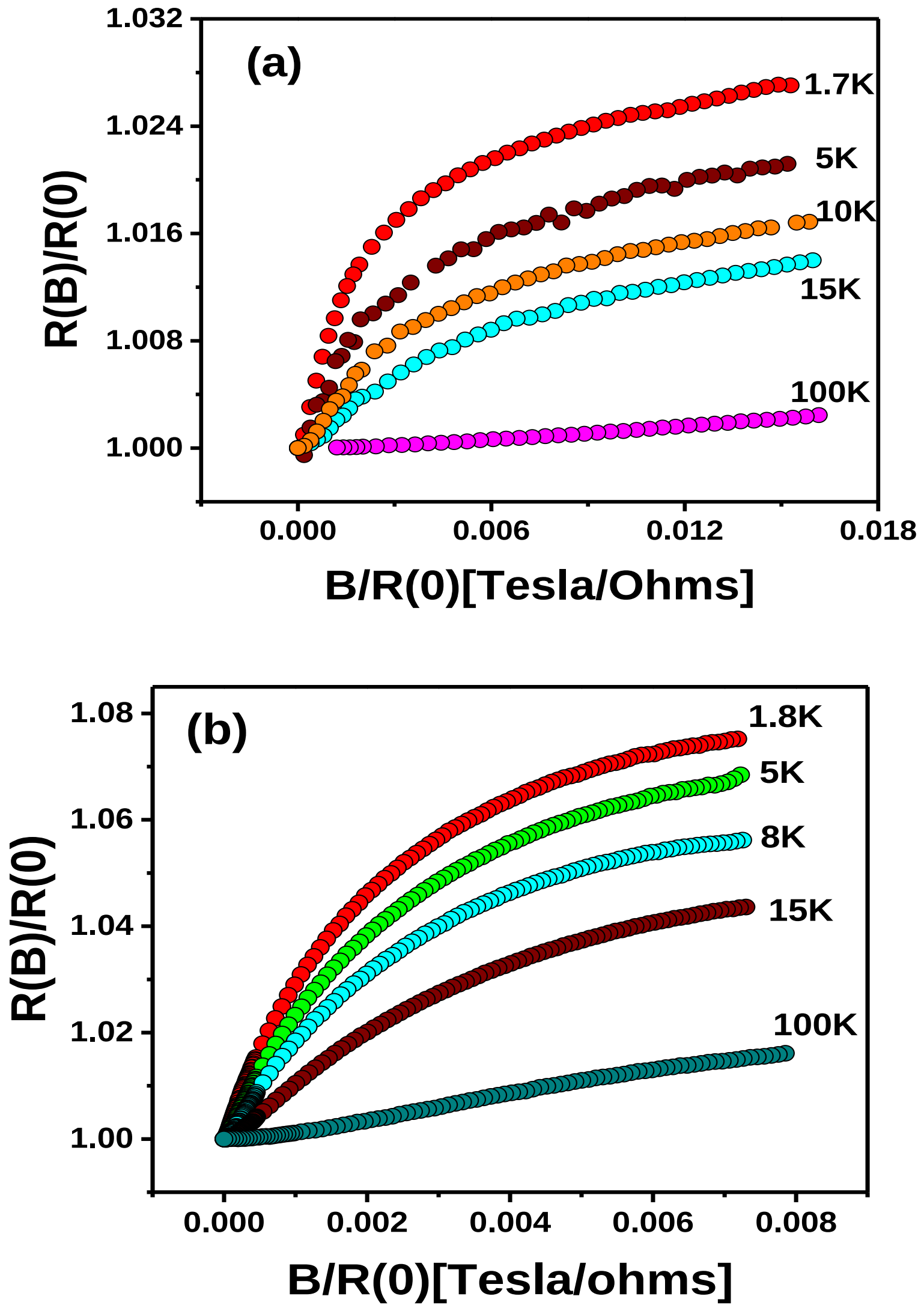

Fig. 2(a) Kohler's plot of BST thin film. (b) Kohler's plot of BSTS thin film. The curves belonging to different temperatures do not fall in a unique curve thus enunciating the violation of Kohler's rule in both the samples.

Since, it has been argued earlier regarding the nature of R-T and MR are intrinsically connected, a similar parallel conduction behavior in the magneto conductivity is expected. It has been previously shown that the inclusion of a quadratic magnetic field term along with HLN term results in linear B dependence[32,33]. This model is represented as an addition of a logarithmic field dependent term due to WAL and a quadratic ($B^2$) term due to the Lorentz force on carriers. This modified equation is a manifestation of parallel conduction between the surface states and bulk states. The parallel conduction model can be written as:

$$G_{Total}\,(T/B) = G_{Surface}\,(T/B) + G_{Bulk}\,(T/B) \qquad (2)$$

In the framework of parallel channel conduction model, temperature dependence of conductance is written as:

$$G_{Total}\ (T) = 1/\ (A+C*T) + 1/D*\exp\ (\Delta/T) \tag{3}$$

Where A is the temperature independent contribution from residual impurity or disorder, C denotes the temperature dependent electron-phonon coupling term. Δ is the energy gap between the impurity band and the bottom of the conduction band. The G-T (conductance vs. temperature) behavior of our BST sample is shown in Fig. 3(b). It is fitted to the parallel conduction model and good fit is obtained[6,10]. The value of the parameters after a perfect fit is given in the supplementary section and is compared with that of BSTS sample.

The magnetoconductance (MC) plots of the BST sample are shown in Figure 3(c) along with the fitted curves. The fitting is done to the modified HLN equation given below. The HLN equation is given by[34]:

$$\Delta G(B) = \frac{e^2}{2\pi h}\left[\psi\left(\frac{B_\varphi}{B}+\frac{1}{2}\right) - \ln\left(\frac{B_\varphi}{B}\right)\right] - \frac{e^2}{\pi h}[\psi\left(\frac{B_{SO}+B_e}{B}+\frac{1}{2}\right) - \ln(\frac{B_{SO}+B_e}{B}+\frac{1}{2})] + \frac{3e^2}{2\pi h}[\psi\left(\frac{\left(\frac{4}{3}\right)B_{SO}+B_\varphi}{B}+\frac{1}{2}\right) - \ln\left[\left(\frac{\left(\frac{4}{3}\right)B_{SO}+B_\varphi}{B}\right)\right] \tag{4}$$

Where, $B_{so}$ ,$B_e$ and $B_\varphi$ are magnetic fields corresponding to spin orbit, mean free and phase coherence lengths. In the high field limit these assumptions: $B \ll B_e$ and $B \ll B_{so}$ do not hold. Moreover, bulk effects start playing an important role in the magnetoconductivity. Inclusion of $B_{so}$, $B_e$ and bulk states contribution leads to the following equation in the high field regime[32]:

$$\Delta G\ (B) = -\frac{\alpha e^2}{2\pi h}[\psi\left(\frac{B_\phi}{B}+\frac{1}{2}\right) - \ln\left(\frac{B_\phi}{B}\right)] + \beta B^2 \tag{5}$$

Where, α denotes the number of channels and β denotes the coefficient of the quadratic term. β comprises of the classical contribution to MC and also the contributions of $B_{SO}\ and\ B_e$ in high field limit[32].

Equation (5) is the representation of parallel conduction model in terms of field dependence. Whereas the fitting of the MC using modified HLN equation (eqn. 5) for BST is shown in figure 3(c), the same for BSTS thin film is shown in figure 4(b). The excellent agreement between the data and the fitted curve supports the argument of parallel channel conduction model. Zhang et al. had earlier stated that LMR is an associated effect of WAL itself[16]. The full HLN equation had been employed to fit their conductance data for the entire field range. Their argument lies in the fact that at higher magnetic field the assumptions of simplified HLN equation does not hold ($B \ll B_e$ and $B \ll B_\varphi$). No additional classical quadratic terms had been added in their HLN fit. The results showed were only up to 30K where the phase coherence length is appreciably large (~ 30 nm). However, when higher temperatures were reached, due to phonon scattering the phase coherence length becomes small and the need to add the classical quadratic terms seems imperative. Moreover, at higher temperatures, the bulk activated behavior dominates the transport properties as it is evident from the present R-T curve. We argue that at higher temperatures, classical contributions (quadratic $B^2$) along with spin–orbit scattering fields ($B_{SO}$) and the field corresponding to the mean free path length ($B_e$) dictate the magnetoresistance properties of the sample.

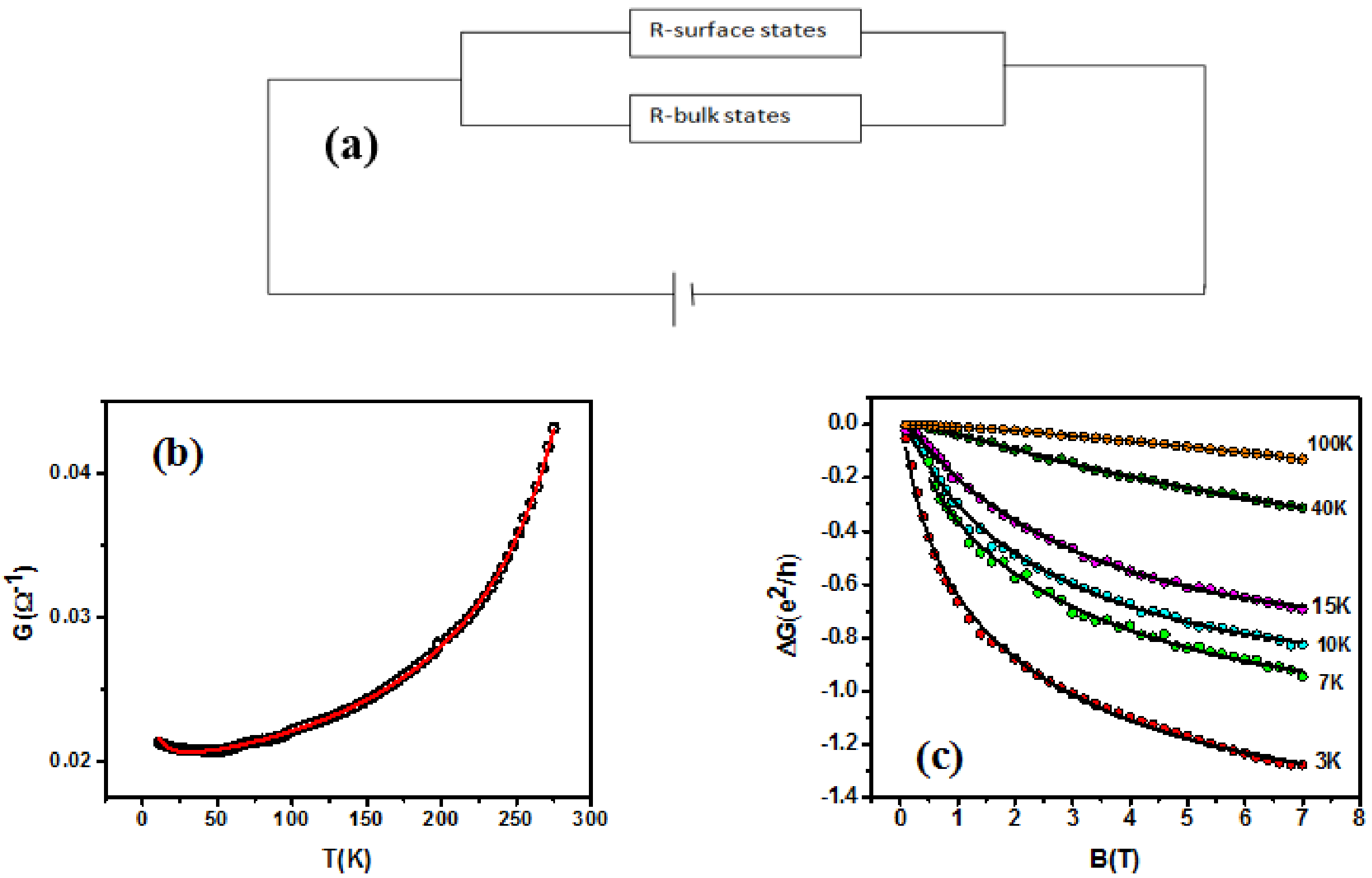

*FIG. 3. (a) Schematic of parallel conduction model. The current will follow the path which is least resistive of the two. (b) Conductance vs. temperature plot of BST thin film. The red curve is the fit using parallel channel conduction model of surface and bulk states. (c) The change in conductance fitted with modified HLN equation (equation 5) which again demonstrates contribution of quantum and classical terms to conductance.*

The field at which the MR behavior changes from logarithmic to linear is called the cross-over field. This cross over field can be seen as the field at which the contributions of the quadratic field term compensates the logarithmic contributions to the magneto-conductivity. A decreasing trend of the cross over field with increasing temperature in the sample was observed (Figure 1(b) and 4(a)). The cross-over can be visualized as a competition between the classical contribution to magnetoresistance and its quantum counterpart. Later, the classical effects compensates the quantum contribution to magnetoresistance and results in a linear MR. Although $B_{SO}$ $and$ $B_e$ gives the quadratic contributions in the high field limit but their contribution is feeble in comparison to the classical contributions arising from the bulk bands[32]. With increasing temperature $l_\varphi$ (phase coherence length) decreases, thus the "quantum-ness" of the system diminishes and classical contributions to the LMR start dominating. Therefore, the cross-over field should decrease as it moves towards higher temperature. This results in contrary to that observed in other Dirac systems where an enhancement of $B_{CR}$ is observed with increasing temperature[35].

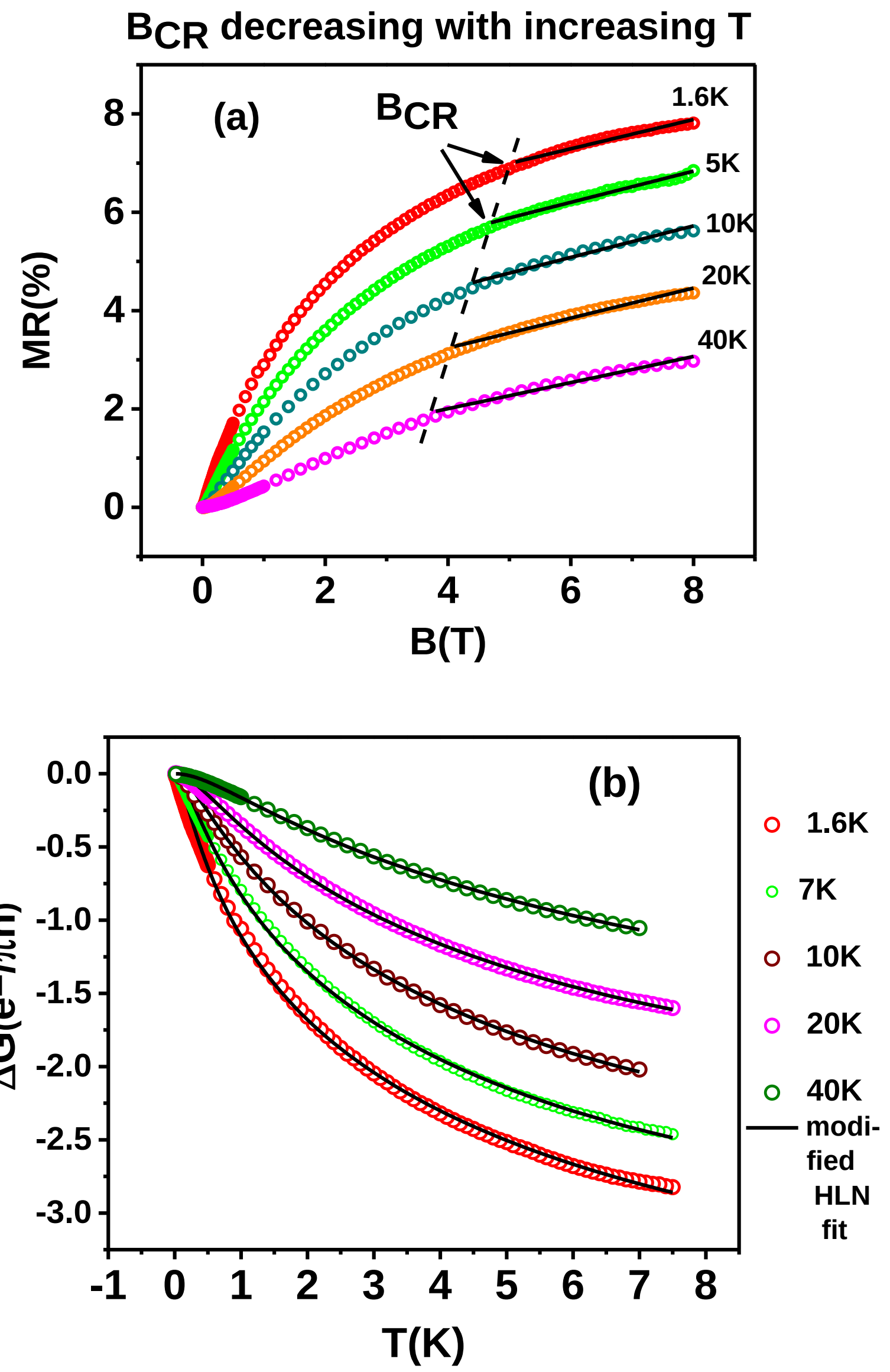

*FIG. **4** (a) MR of BSTS sample for various temperatures. The black straight line fit for each temperature depicts the LMR regime and the cross-over field decreases as we increase the temperature. (b) Modified HLN fit of BSTS data at different temperatures.*

| Temperature (K) | α (**BST**) | $B_{CR}$ (**BST**) | α (**BSTS**) | $B_{CR}$ (**BSTS**) |
|---|---|---|---|---|
| 1.6 | 0.96 | 4.01 | 1.02 | 4.25 |
| 3 | 0.93 | 3.92 | 1.07 | 4.10 |
| 7 | 0.97 | 3.80 | 1.05 | 3.95 |
| 10 | 1.05 | 3.58 | 1.05 | 3.70 |
| 20 | 0.81 | 3.20 | 0.97 | 3.50 |
| 40 | 0.45 | 2.50 | 0.54 | 3.10 |
| 100 | 0.35 | 2.00 | 0.48 | 2.75 |

Table **1**. Conductance data fitted to “modified HLN” equation (4) and parameters obtained are shown in table as a function of temperature. Black is for BST sample and red is for BSTS sample.

Due to unintentional defect formation in bulk during film growth even insulating thick samples (200 nm) can exhibit indirect surface to bulk coupling[36,37].In an earlier work the value of “α”was used phenomenologically to distinguish the different coupling regimes[11].This surface to bulk coupling leaves its signature on the LMR behavior too. The present study proposes that $B_{CR}$ is a better tool to understand the strength of surface to bulk coupling in a sample as formation of Two Dimensional Electron Gas (2DEG) can lead to misleading values of α [see supplementary]. The cross over field physically captures the resultant of bulk and surface contributions to magnetoconductivity. At a particular temperature the sample with a strong surface to bulk coupling will have a lower $B_{CR}$ in comparison to a sample having weak surface to bulk coupling. A weak coupling implies a high bulk insulating behavior which in turn lowers the quadratic bulk contribution to MC. Thus it would require a comparatively high magnetic field value to achieve this crossover. A strong coupling, on the contrary, would result in lowering of the crossover field value (see crossover field in Table 1). BSTS is a better bulk insulator than BST as evident from the Hall data [supplementary]. Surface to bulk coupling is accordingly lesser in BSTS and that results in lowering of the cross over field (Table 1). Thus, in conclusion the cross over field is an indication of the extent of surface to bulk coupling.

Although our main focus in this paper is to investigate the LMR in TI films, but in low fields these films show pronounced WAL character, despite the presence of heavy disorder in the form of grain boundaries. Recently we became aware of two reports with contrasting experimental observations. In one report (ref.39) author claims that grain boundary disorder in the $Bi_2Se_3$ thin films drives the surface Dirac states into strongly insulating regime implying that the surface states are not immune to strong localization. Whereas in second (ref.38) it is found that the Dirac states remain immune to this type of disorder at temperature down to 0.3K [38, 39]. We also did not get any sign of strong localization in our BST and BSTS films down to 1.6K; instead the surface electrons remain topologically protected and display enhanced metallicity as the temperature is reduced. Our results support the experimental findings of ref.38 and the surface electrons remain topologically protected and display enhanced metallicity as the temperature is reduced. These polycrystalline films, especially BSTS which is an alloyed solution of Bi2Se3 and Sb2Te3, displays more bulk insulating character and pronounced WAL, contains heavy grain boundary disorder in comparison to the MBE grown single phase films (see supplementary information).

**CONCLUSION:**

LMR on bulk insulating TI thin films was studied and the role of both surface and bulk states was invoked to explain the observed phenomena at high field values. With increasing temperature the phase coherence length decreases and the dominance of bulk states soars over the surface state contribution. The LMR originates due to the presence of both surface states and the bulk states and their combined contribution to the magnetoresistance. The magneto-conductance data is fitted with a modified HLN equation which is an illustration of the parallel conduction model. The overall result is lowering of cross-over field where LMR sets in with increasing temperature. This temperature dependent cross-over field behavior is associated with surface to bulk coupling in TI.

**SUPPLEMENTARY MATERIAL:** Information regarding thin film growth, characterization, measurement and relevant explanation is provided in the supplementary section.

**ACKNOWLEDGEMENTS:** Authors would like to thank Ministry of Human Resource Development for generous funding and IISER Kolkata for instrument facilities. RK Gopal would like to thank Suman Sarkar for valuable discussions and inputs. SS and JS would like to thank University Grants Commission (UGC) for fellowship and contingency grants.

# Authors have contributed equally in this work.

# Supplementary Material

## Linear magnetoresistance and surface to bulk coupling in topological insulator thin films

**Sourabh Singh[1#], R.K. Gopal[1#], Jit Sarkar[1], Atul Pandey[1], Bhavesh G. Patel[2] and Chiranjib Mitra[1]**

***[1]Indian Institute of Science Education and Research Kolkata, Mohanpur,741246***

***India***

***[2]Institute of Plasma Research, Bhat, Gandhinagar,382428, India***

### Sample preparation and characterization:

Thin films of $Bi_2Se_2Te$ (BST) and $BiSbTeSe_{1.6}$ (BSTS) were deposited by pulsed laser deposition (PLD) technique. Targets were ablated by KrF laser (248 nm) and deposited on Si (100) undoped substrates in a flowing Argon environment. The substrates were thoroughly cleaned and ultra-sonicated sequentially with Acetone, Isopropanol and DI water. Quality of the prepared thin films depend on factors like Argon pressure, substrate temperature, laser fluence, annealing temperature and distance between target and substrate. These parameters were optimized to obtain the best quality thin films.

Prepared thin films were characterized by X-ray diffraction (XRD) and Raman spectroscopy. The surface of a TI is sensitive to external environmental exposure which cause formation of unwanted trivial surface states and hamper to get the intrinsic surface state properties of TI films. While doing optical lithography the upper surface of the film is subjected to chemicals which in turn can act as a detrimental factor in the determination of surface transport properties. Therefore, we prepared these films by shadow mask technique where a standard 6 terminal Hall bar mask of known dimensions was placed on top of a substrate(see figure S1 (c)).

s

(a

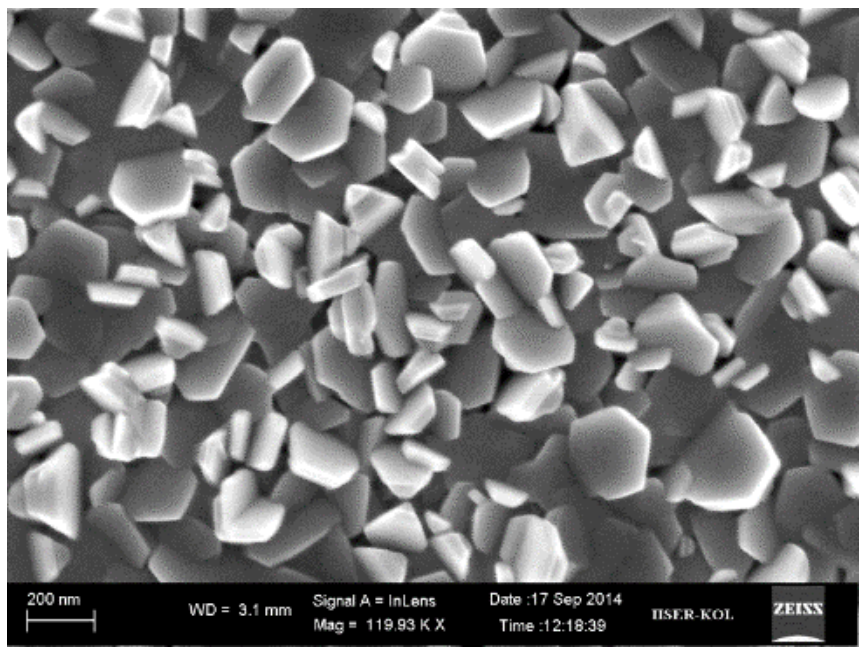

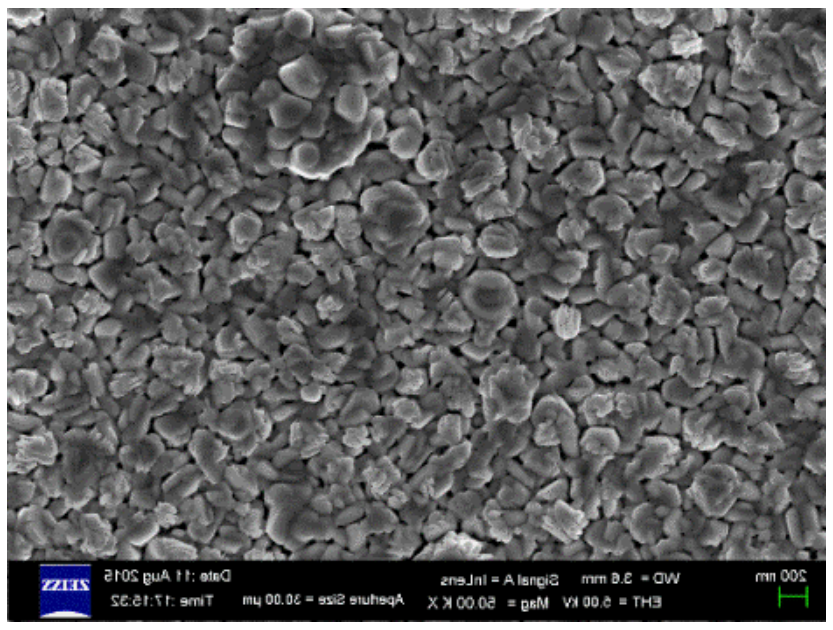

(b)

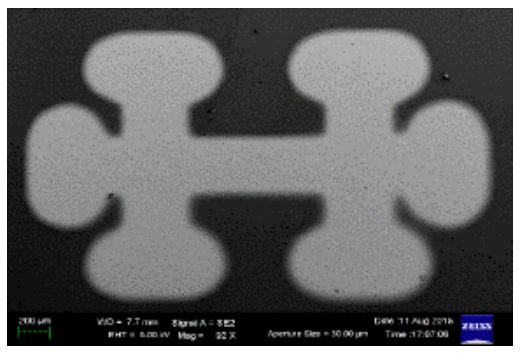

(c)

Figure S**1**. (a) SEM image of BST thin film (b) SEM image of BSTS thin film and (c) SEM image of film prepared with shadow mask technique. A TI film grown using six terminal Hall bar mask is shown in the figure (c).

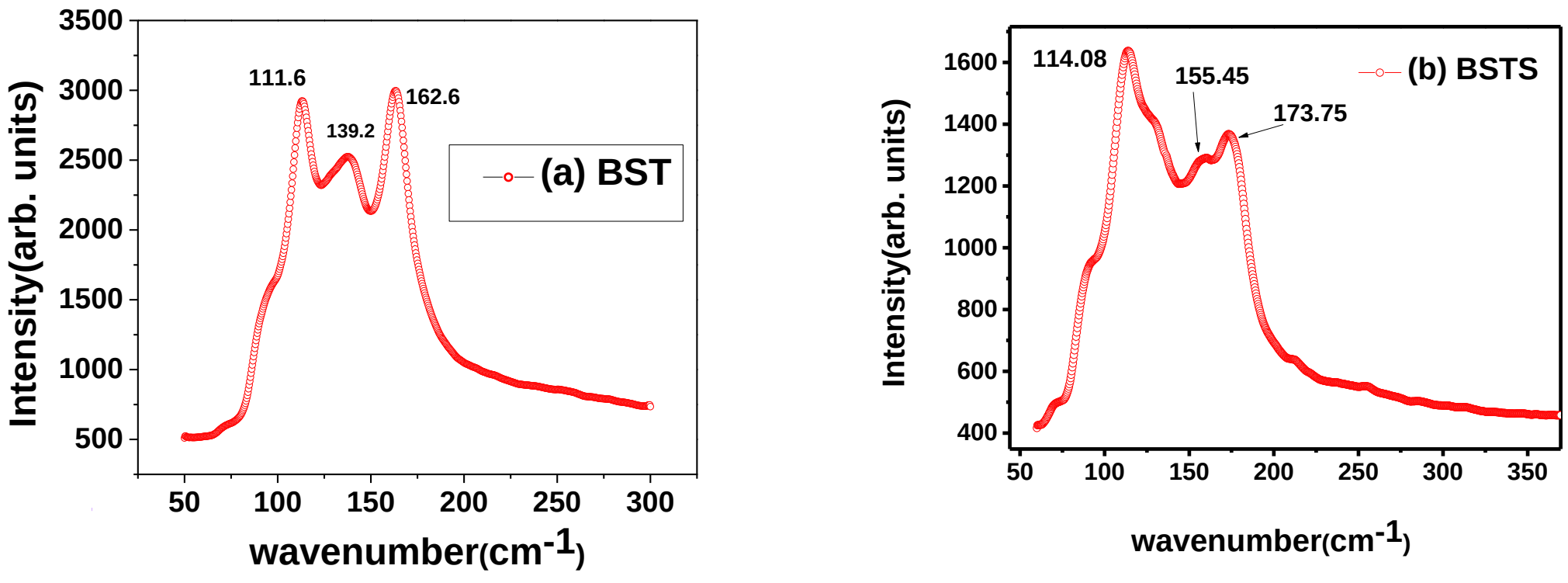

Figure S**2**. (a) Raman spectrum of BST thin film. The peak positions at 111.6 $cm^{-1}$ and 139.2 $cm^{-1}$ correspond to doubly degenerate $E^2_g$ mode and 162.6 $cm^{-1}$ belong to $A^2_{1g}$ mode. (b) Raman spectrum of BSTS film with the first peak corresponds to $E^2_g$ mode at 114.08 $cm^{-1}$, two other peaks correspond to $A^2_{1g}$ mode at 155.45 and 173.75 $cm^{-1}$.

Raman spectrum of the BST and BSTS thin films is shown in the fig.S2, which is consistent with the layered chalcogen ordering in the tetradymite crystal structure. The Raman peaks in BSTS match with the previous reports[1,2].

Magnetoresistance (MR) and Hall measurements were taken together using two separate nano voltmeters and a standard lock in technique in cryogen free Cryogenic system from 1.6K-300K and 0-8T. The magnetoconductance is calculated from the resistance using the following formula; $\Delta G (B) = [R (0)-R (B)]/[R (0)]^2$ and magnetoconductivity is obtained using $\Delta\sigma (B) = \Delta G (l/w)$. Where, "l" and "w" are the distance between the two leads and the width of the leads, respectively.

**Linear Magneto-resistance (LMR):**

The LMR model proposed by Abrikosov also known as the quantum LMR model is valid only for samples having linear dispersion relation and are in the extreme quantum limit[3,4]. The system is said to be in the extreme quantum limit when only the first Landau level is occupied and rest of the levels are vacant. When the Fermi energy of the system is smaller than the difference between the energy of the lowest Landau level and first excited level then the system is said to being the extreme quantum limit.

$$E_F, k_BT < E_1 - E_0,$$

Where, $E_F$ stands for Fermi energy, $k_B$ is the Boltzmann constant; $E_0$ and $E_1$ are energy levels corresponding to the lowest and first excited Landau levels respectively.

TI samples have a large Fermi surface, hence carrier density in these samples are large. Hence very high magnetic field is required to obtain the extreme quantum limit. To achieve the extreme quantum limit the following conditions should be satisfied by the magnetic field and the temperature:

$$n \ll \left(\frac{eB}{\hbar}\right)^{3/2},\ T < v\sqrt{eB\hbar/}\ k_B \qquad (1)$$

Where, carrier density (n) in our case is of the order of $10^{18}$ $cm^{-3}$ which is too high for fulfilling equation (1). The magnetic field required to fulfill this condition is very high as compared to the maximum magnetic field used in this experiment (8T). "e" stands for electron charge, $\hbar$ is the reduced Planck's constant, $k_B$ is the Boltzmann constant and v depicts velocity of the electron. Important observation of quantum LMR model is that LMR should increase as we increase the temperature but in our sample the LMR is decreasing with increasing temperature. Therefore, we can conclude that the LMR observed in our samples is not due to quantum LMR as proposed by Abrikosov.

Parish and Littlewood (PL) came up with a classical model for LMR[5,6]. They proposed that inhomogeneity in disordered conductors give rise to LMR. This implies that LMR is governed by carrier mobility. Mobility can be used as a measure to understand the amount of disorder in a sample. According to this model, there should be a correlation between the slope of the MR and mobility of the sample[7].

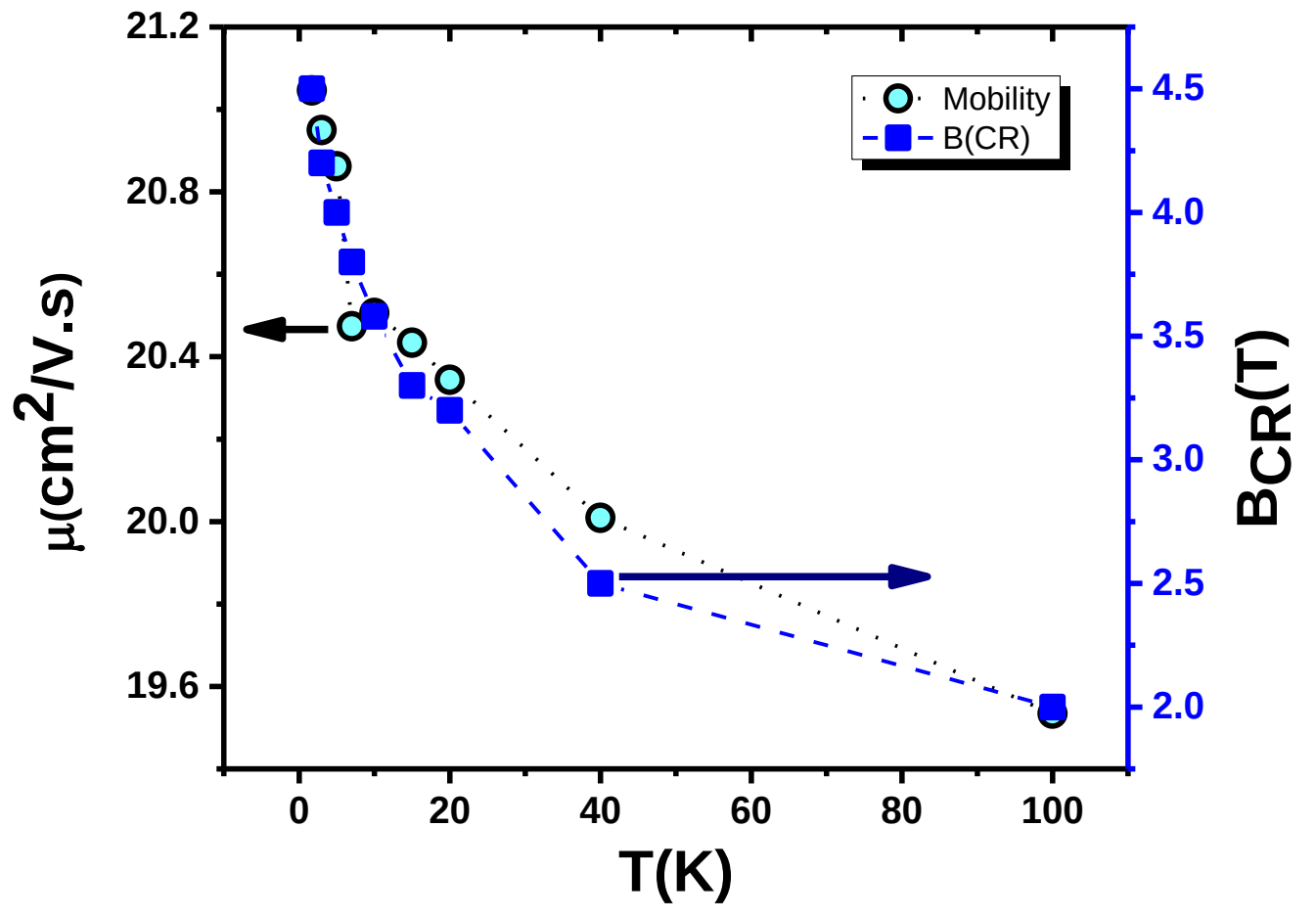

*FIG.S**3**. Double axis plot as a function of temperature for BST sample. The left axis correspond to the Hall mobility whereas the right axis is cross over field.*

The PL model further predicts that with increasing temperature the cross-over field ($B_{CR}$)should increase. This is because, increasing temperature results in reduction of Hall mobility and the cross-over field is inversely proportional to the mobility. But from Fig.S3 we can see that our data depicts a different story altogether. As the temperature is gradually increased there is a reduction in the cross-over field value. It has been previously reported that decreasing the film thickness beyond 6nm, results in the vanishing of LMR. This observation is a clear-cut demonstration that the surface state plays a pivotal role in the occurrence of LMR. Since upon reducing the thickness of the film below 6nm, a gap opens at the Dirac point, and the system becomes a trivial insulator or a bad conductor with no topological protection. PL model, on the other hand, does not mention the requirement of linear dispersion or presence of surface states for the occurrence of LMR. Thus we can conclude that the LMR in our sample cannot be explained by the classical PL model.

To obtain the cross-over field we calculated dMR/dB (derivative of MR with respect to field) and plotted it as a function of B (magnetic field)[8]as shown in figure S4.. The point at which there is a change in slope in the curve is the cross-over field. The plot depicts a reduced saturating slope enunciating the fact that beyond cross-over field the MR is linearly dependent on B. This crossover field has different physical meaning in different interpretations of LMR. In the Abrikosov's quantum LMR model this corresponds to the field at which all the electrons fall in the first Landau level. In the PL classical LMR model, the cross-over field is seen as the measure of average mobility.

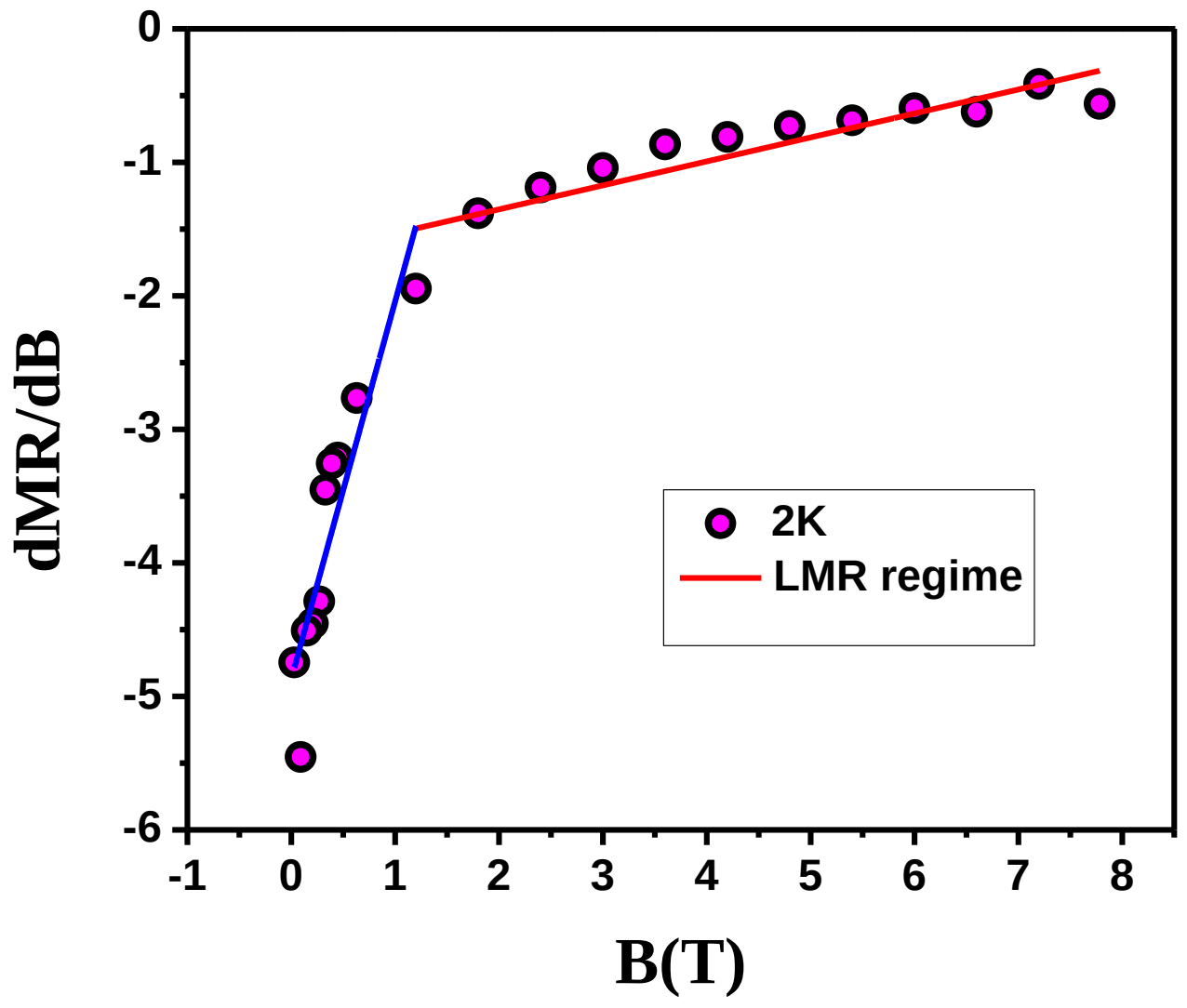

Figure S**4**. Derivative of MR with respect to magnetic field plotted as a function of magnetic field (BST sample). The change in slope depicts the onset of linear MR (red solid line) and the field value at which this happens is the cross-over field.

Figure S**5** validates the point that increasing temperature lowers the phase coherence length and thus it is easier for classical terms to compete with the quantum contribution to magnetoresistance. Thus the phase coherence length which is a measure of the “quantum-ness” of the system is inversely correlated with the cross-over field.

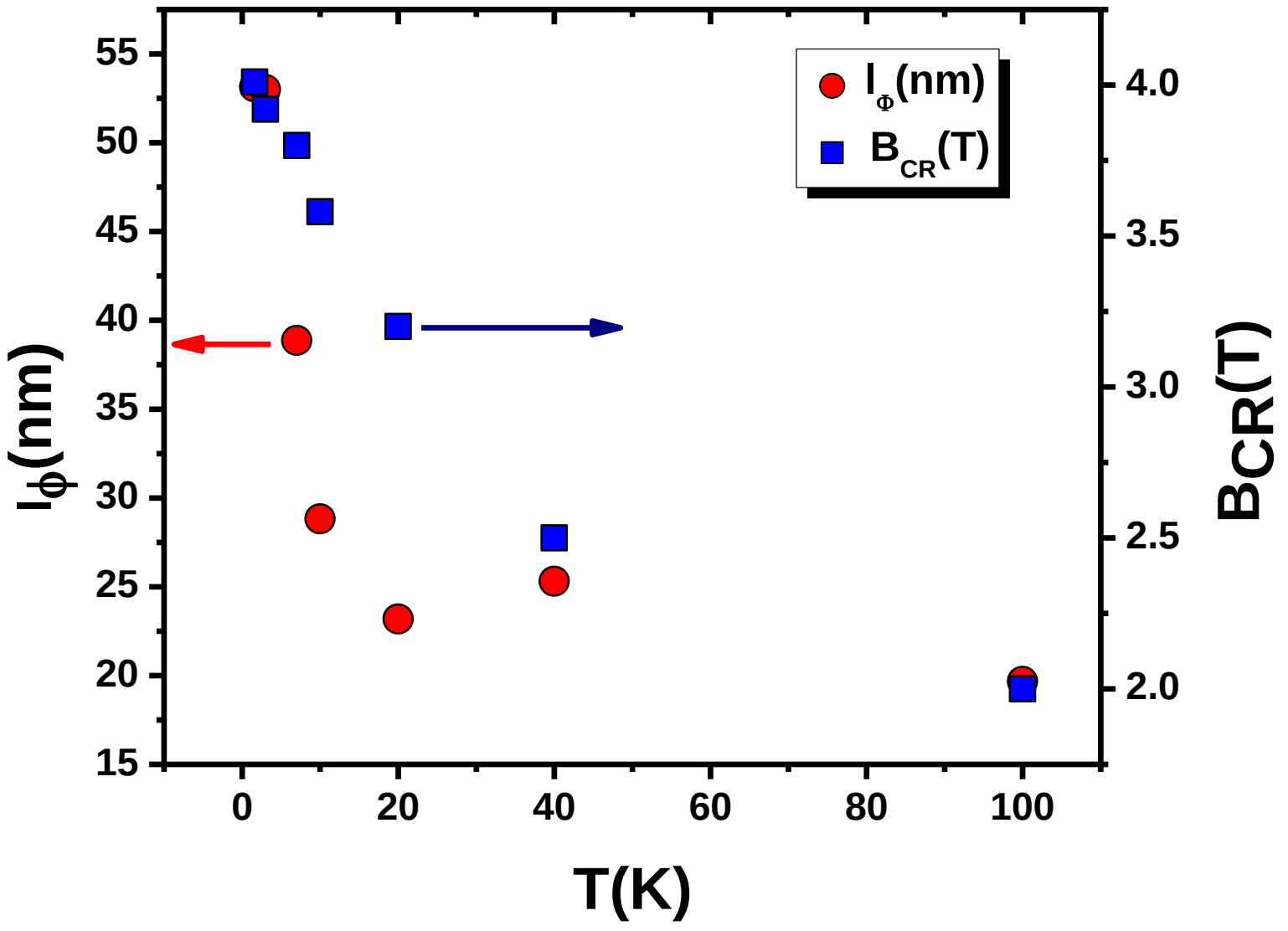

*FIG.S**5**. Variation of cross over field with temperature for the BST sample. A similar trend is observed for the BSTS sample. Increasing the temperature lowers the cross-over field value thus exhibiting the dominance of bulk conduction at higher temperature.*

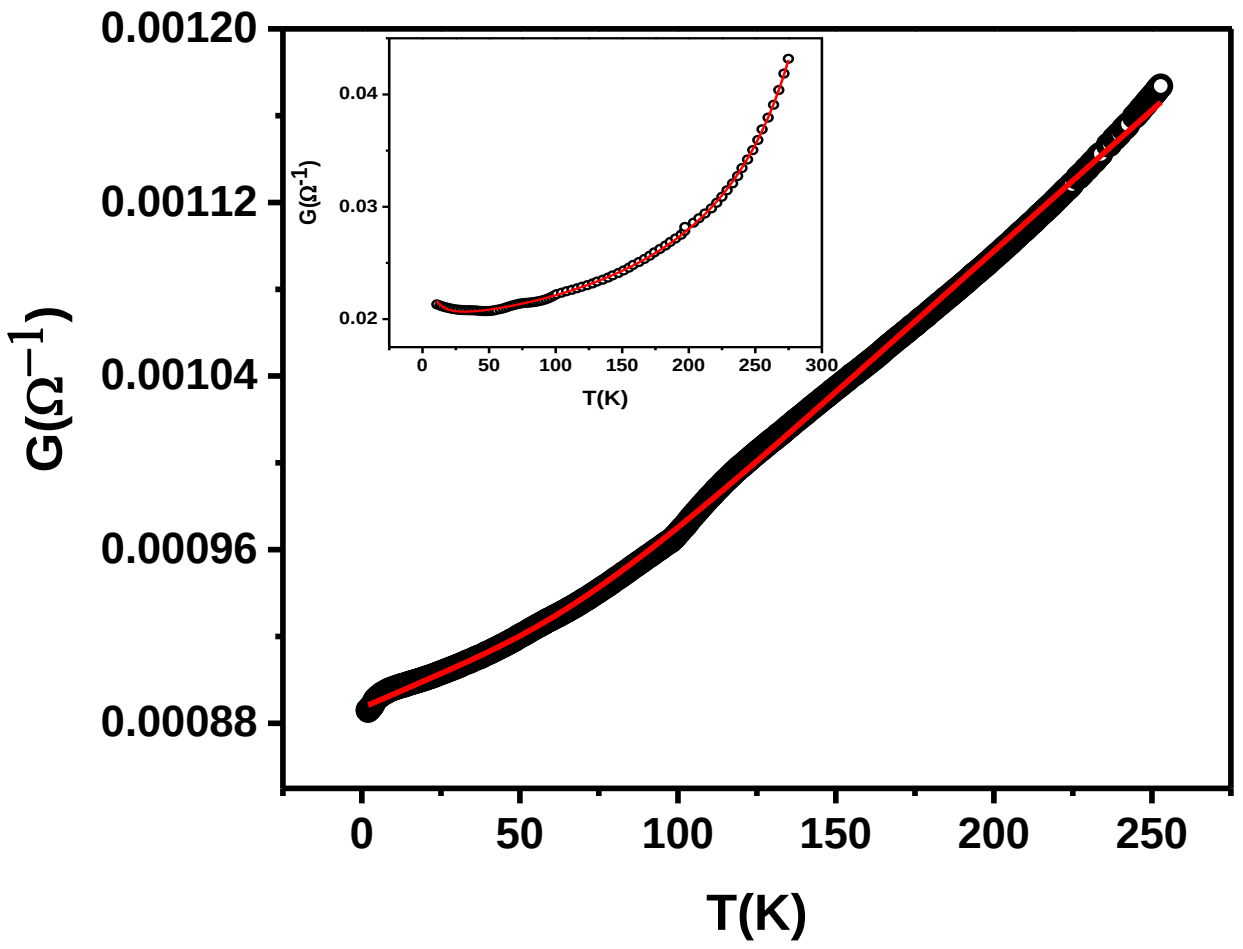

*Fig. S**6**. Temperature dependence of conductance of BSTS sample. The red solid line is the fit using parallel conduction model using equation (2). The inset shows the corresponding graph of BST sample.*

| SAMPLE | A | C | Δ (meV) | D |
|---|---|---|---|---|
| BST | 52.52 | 0.092 | 23.97 | 1645.31 |
| BSTS | 1127.19 | 0.78 | 114.7 | 3739.09 |

Table S1. Comparison of the fitted parameters from the R-T of the two TI films. The activation gap Δ indicates the position of the impurity band corresponding to the conduction band bottom. Thus, BSTS sample is a better insulator than BST thin film.

**The problem of using α as a tool to parameterize surface to bulk coupling:**

Using α as a phenomenological tool to quantify surface to bulk coupling hits a major roadblock owing to the existence of two dimensional electron gas (2DEG) on the surface. This 2DEG participates in the 2D electron conduction along with the non- trivial Dirac fermions. The 2DEG is formed at the surface of TI's as a result of electron doping when TI samples are exposed to the external environment[9,10]. Formation of 2DEG results in downward bending of bands near surface making a potential gradient which is asymmetric in nature hence it is of Rashba type. This trivial 2DEG is confined in this triangular like potential and forms separate parabolic bands coexisting with nontrivial topological states, which may split if confinement is stronger or exposure to the external environment is larger. This is captured schematically in Fig. S7 (a). This 2DEG could be localized completely at low temperatures by the disorder and grain boundaries present in PLD grown thin films. Therefore, there is no general consensus of the value of α obtained from fitting the HLN equation. Hence a new tool to understand and quantify surface to bulk coupling was called for and we propose that the cross-over field value is a better measure.

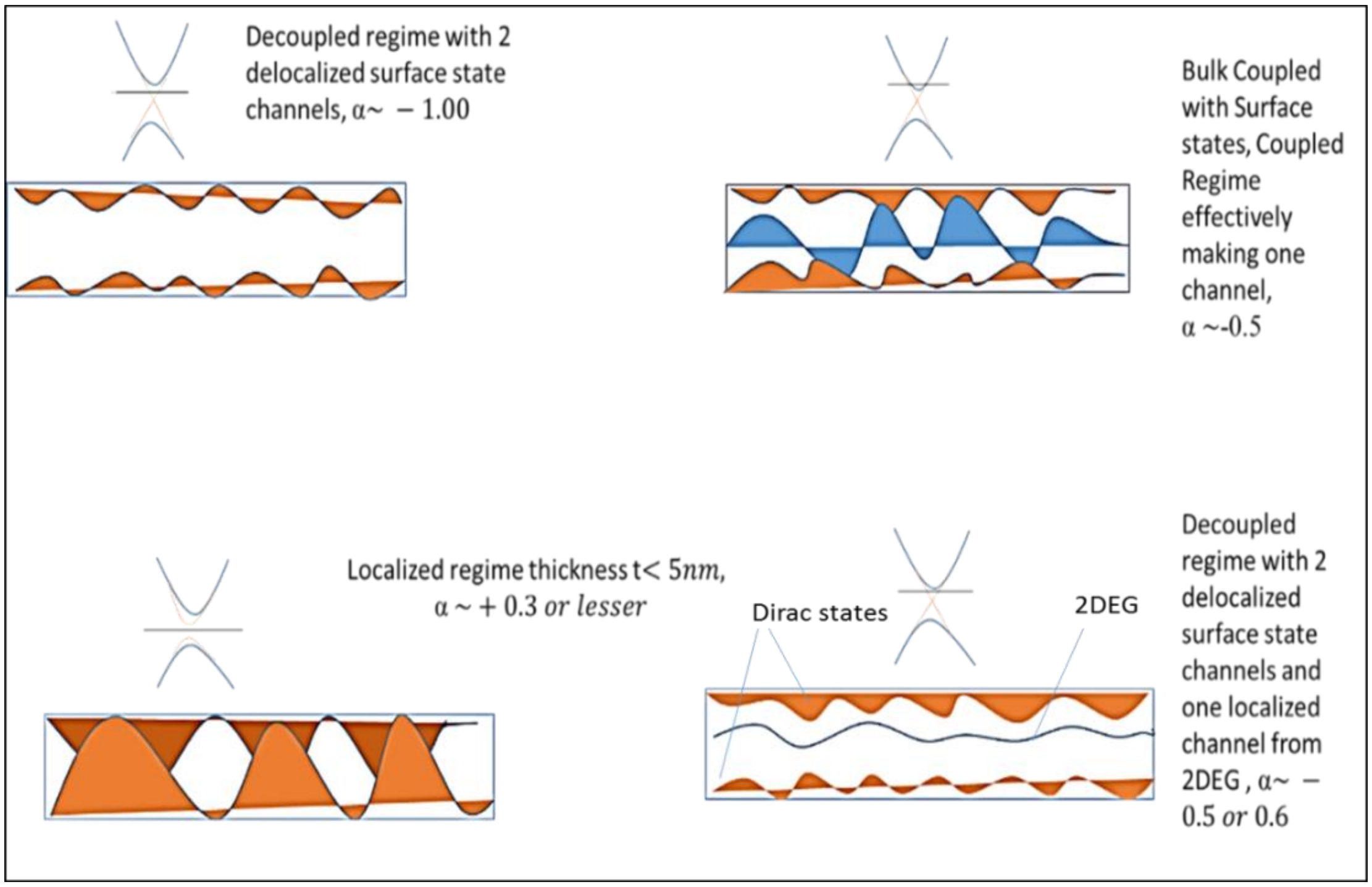

FIG.S**7.** Schematic explanation of coupled and decoupled transport of surface states in different regime. **(a)** Decoupled transport of by two surface channels in bulk insulating sample providing value of the coefficient α = (-½)+ (-½) =-1.**(b)** Metallic thin films where bulk and surface channels are coherently coupled, effectively resulting in one channel α = -½ . **(c)** Strongly localized regime when the thickness of the film is below critical thickness t <5nm. **(d)** Decoupled transport in a bulk insulating film when exposed to the external environment. Due to the downward band bending a trivial 2DEG is formed beneath the upper surface. Apart from delocalized Dirac fermions on two surfaces this confined 2DEG gets localized at low temperature and the value of α is the result of mixed contribution from weak localization and weak antilocalization phenomena.